\documentclass[cameraready]{Interspeech}
 
\usepackage{algorithm}
\usepackage{algpseudocode}
\usepackage{booktabs,multirow,graphicx}
\title{Elastic Time: Dynamic Frame Rate Bottlenecks for Neural Audio Coding}

\author[affiliation={1}, orcid=0009-0009-3127-0111]{Dimitrios}{Bralios}
\author[affiliation={2}, orcid=0009-0006-8623-4175]{Paris}{Smaragdis}
\author[affiliation={1}, orcid=0000-0003-3513-8328]{Minje}{Kim}
\address{
    $^1$ University of Illinois Urbana-Champaign, Urbana, IL, USA \\
    $^2$ Massachusetts Institute of Technology, Cambridge, MA, USA
}

\email{dbralios@illinois.edu, paris@mit.edu, minje@illinois.edu}

\keywords{neural audio autoencoders, dynamic frame-rate, adaptive rate}

\usepackage{comment}

\begin{document}

\maketitle

% Dynamic Frame Rate Re-Bottleneck

\begin{abstract}

Neural audio autoencoders have become a core component of compression, feature extraction, and generation. However, while existing systems support variable bitrate, the vast majority of models still operate at a fixed latent frame-rate, allocating equal temporal budget to regions with very different information density, which can result in unnecessarily long sequences. We introduce Elastic Time, a dynamic frame-rate bottleneck that converts fixed-frame-rate autoencoders to dynamic ones. Our method learns a lightweight latent predictor used to decide which frames can be skipped and later reconstructed, enabling efficient greedy boundary selection at inference. Experiments show our method enables deployment-time rate control while improving efficiency-quality tradeoffs relative to baselines. Overall, we provide a flexible mechanism for adjusting temporal resolution in audio autoencoders, potentially facilitating more efficient downstream modeling for generation and long-context tasks.

    % Manuscripts submitted to Interspeech 2026 must use this document as both an instruction set and as a template. Do not use a past paper as a template. Always start from a fresh copy, and read it all before replacing the content with your own. The main changes with respect to previous years' instructions are \blue{highlighted in blue}.

    % Before submitting, check that your manuscript conforms to this template. If it does not, it may be rejected. Do not be tempted to adjust the format! Instead, edit your content to fit the allowed space. \blue{The maximum number of manuscript pages is 6 for regular papers and 10 for long papers. For regular papers, pages 5 and 6, and for long papers, pages 9 and 10, are reserved exclusively for acknowledgments, disclosures of the use of generative AI tools, and references, which may begin on an earlier page if space permits.}

    % The abstract is limited to 1000 characters. The one in your manuscript and the one entered in the submission form must be identical. \blue{In the submission form, no \LaTeX{} code is allowed.} Do not use citations in the abstract: the abstract booklet will not include a bibliography. Index terms appear immediately below the abstract.
\end{abstract}

\section{Introduction}
\label{sec:introduction}

% Neural audio codecs and autoencoders
Advances in neural audio codecs and autoencoders have enabled the compression of audio waveforms into compact latent codes, significantly improving storage and transmission efficiency while maintaining perceptual quality and content fidelity \cite{defossez2023high, zeghidour2021soundstream, kumar2023high, li2025spectrostream, evans2025stable}. 
Beyond compression, these learned representations have become a core layer for generative audio modeling. Discrete codecs yield token sequences that naturally support language modeling approaches for audio generation, including autoregressive next-token prediction \cite{borsos2023audiolm, kreukaudiogen, copet2023simple, defossez2024moshi} and parallel token prediction \cite{borsos2023soundstorm}. Continuous latents \cite{evans2025stable, wang2025back, braun2025salad}, in contrast, are commonly used as the modeling space for diffusion-based generative models \cite{evans2025stable, hai25_interspeech}, though more recently have also been the predictands of an autoregressive sequence model~\cite{rouard2026continuous}.
For continuous representations, since quantization-based compression is impossible, reducing data rate by lowering dimensionality or frame-rate is more critical than using discrete tokens.

Unlike compression, where overall bitrate and reconstruction quality are the primary evaluation metrics, generative models are often bottlenecked by compute and memory, which scale with the effective latent sequence length. Language-modeling approaches operating 
on residual vector quantization (RVQ)~\cite{zeghidour2021soundstream} codes can reduce the dependence on the number of codebook levels via interleaved codebook patterns \cite{copet2023simple}, depth-wise prediction heads \cite{defossez2024moshi,rouard2026continuous}, or parallel prediction of finer levels, but the dominant cost remains driven by the token rate and the context length required for generation. A similar dependence holds for latent diffusion \cite{evans2025stable,hai25_interspeech}. While reducing latent dimensionality (through the channel count) can facilitate modeling \cite{zheng2026diffusion}, attention-based diffusion backbones such as diffusion transformers incur costs that grow with the latent sequence length. Consequently, latent sequence length is a key design choice, especially for long-context generation~\cite{evans2024long}.

While variable-bitrate neural codecs and autoencoders exist, most prior work adjusts capacity per latent frame.
For instance, through structured dropout during training, we can construct scalable models that work with a varying number of residual codebooks~\cite{zeghidour2021soundstream}. 
Additionally, works like VRVQ~\cite{chae2025variable} enable input-adaptive variable rate by automatically picking the number of codebooks used at each time step. 
In contrast, work that enables scalable and/or adaptive latent frame-rate is comparatively sparse~\cite{zhang2025unlocking, li2026flexicodec, della2026beyond, wang2025codecslime, zheng2025say, chae2025variable}. Most existing work employs external semantic information to guide decisions.
TFC~\cite{zhang2025unlocking} proposes the use of temporal entropy in the input waveform as a proxy for frame-rate selection decisions. FlexiCodec~\cite{li2026flexicodec} computes similarity between neighboring pretrained ASR features to produce latent segment boundaries. DyCAST~\cite{della2026beyond} uses a pretrained character aligner to provide supervision for chunking. CodecSlime~\cite{wang2025codecslime} allows dynamic frame-rate without external supervision via a dynamic programming algorithm, but requires a two-stage training setup. 

We introduce Elastic Time, a plug-in Re-Bottleneck module~\cite{bralios2025re} that converts a pretrained fixed-frame-rate audio autoencoder into a dynamic-frame-rate model. 
We focus on the finite-segment, offline autoencoding setting, where the full encoded segment is available before temporal rate allocation.
Our core technical contribution is a lightweight predictor trained to predict future latent frames from past context.
%whose resulting predictability is used as a proxy for temporal redundancy: frames that are more accurately predictable can be omitted and approximated from nearby context. 
We use its prediction accuracy as a proxy for temporal redundancy: frames predicted accurately can be omitted and reconstructed from nearby context.
During offline inference, given a user-specified average compression factor, our model adaptively selects segment boundaries in the latent sequence and then reallocates temporal rate budget by solving a resource allocation problem. We propose an efficient greedy algorithm as well as a dynamic programming-based optimization procedure. Crucially, our method requires no external semantic supervision (e.g., foundation model features) during training or inference. Our source code is publicly available\footnote{
\url{https://github.com/dbralios/elastic-time}
}.

Training neural audio autoencoders from scratch is expensive and requires careful tuning of adversarial/perceptual losses. Notably, there is a plethora of pretrained, publicly available fixed-rate models. The Re-Bottleneck framework lets us reuse these pretrained networks and focus on learning a latent-space module that specializes in dynamic temporal allocation while preserving the base model's strong reconstruction ability.

\section{Proposed Method}
\label{sec:method}

\begin{figure}[t]
    \centering
        \centering
        \includegraphics[width=.95 \columnwidth, trim=7.5pt 0pt 0pt 0pt, clip]
        {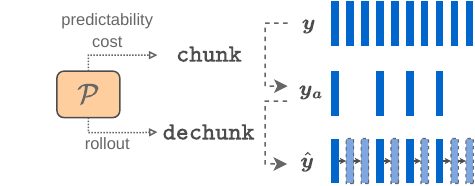}      
        \caption{Chunk and dechunk procedures. Chunking retains anchor frames $\bm{y}_{\bm{a}}$ based on latent predictability. Dechunking uses predictor $\mathcal{P}$ to expand $\bm{y}_{\bm{a}}$ back to full-length sequence $\hat{\bm{y}}$.}
    \label{fig:fig1}
    \vspace{-0.75em}
\end{figure}

\subsection{Overview}

We assume a pretrained autoencoder with encoder $\mathcal{A}_{\texttt{enc}}$ and decoder $\mathcal{A}_{\texttt{dec}}$, which maps a multichannel audio waveform $\bm{x}\in\mathbb{R}^{C_\text{audio}\times L}$ of length $L$ to a latent sequence and reconstructs it back to audio
\begin{equation}
\bm{z} = \mathcal{A}_{\texttt{enc}}(\bm{x}), \quad \bm{z}\in\mathbb{R}^{C\times T},\quad
\hat{\bm{x}} = \mathcal{A}_{\texttt{dec}}(\bm{z}).
\end{equation}
More compression is achieved when $T\ll L$, but, in addition to that, our goal is to turn this fixed latent frame-rate into a dynamic one, controlled at inference time by a per-sample target kept length $N$ (equivalently, a kept fraction $\rho = N/T$). 
To this end, we introduce a Re-Bottleneck module with encoder $\mathcal{R}_\texttt{enc}$ and decoder $\mathcal{R}_\texttt{dec}$, without modifying the number of latent dimensions $C$, that yields
\begin{equation}
\bm{y} = \mathcal{R}_{\texttt{enc}}(\bm{z}), \quad \bm{y}\in\mathbb{R}^{C\times T}
%,\quad\hat{\bm{z}} = \mathcal{R}_{\texttt{dec}}(\hat{\bm{y}}).
\end{equation}
We also introduce a learned latent predictor $\mathcal{P}$ that models latent dynamics.
This enables a \texttt{chunk} procedure that partitions the latent sequence into $N$ variable-length segments according to a predictability criterion, retaining one ``anchor'' frame from each segment.
Then a \texttt{dechunk} procedure produces the full-length reconstructed latent sequence $\hat{\bm{y}}$, using $\mathcal{P}$. We expand on these procedures (visualized in Fig.~\ref{fig:fig1}) in following Sections. Finally, we map back to the autoencoder latent space and reconstruct audio:
\begin{equation}
\hat{\bm{z}} = \mathcal{R}_{\texttt{dec}}(\hat{\bm{y}}), 
\qquad \hat{\bm{x}} = \mathcal{A}_{\texttt{dec}}(\hat{\bm{z}}).
\end{equation}

\subsection{Latent predictor}
\label{subsec:predictor}

We employ a lightweight autoregressive predictor $\mathcal{P}$ that models short-term latent dynamics along the temporal axis. Given the previous predicted frame $\hat{\bm{y}}_t$ and an optional recurrent state $\bm{h}_t$, $\mathcal{P}$ predicts the next latent frame:
\begin{equation}
\hat{\bm{y}}_{t+1}, \bm{h}_{t+1} 
= \mathcal{P}(\hat{\bm{y}}_{t}, \bm{h}_t). 
% \qquad t=0,\dots,T-2.
\label{eq:predictor_step}
\end{equation}
The predictor is used in open-loop mode during dechunking, using its previous predictions as inputs (Eq.~\eqref{eq:dechunk_rollout}). This motivates training objectives explicitly penalizing multi-step rollout error.

\subsection{Chunking}
\label{subsec:chunking}

The chunking procedure returns chunk boundaries, as shown 
\begin{align}
\bm{b}& \leftarrow \texttt{chunk}\left(\bm{y} ; N, \mathcal{P} \right), \\
0 &=b_0 < b_1 < \cdots < b_{N-1} < b_{N} = T,
\end{align}
where $\bm{b} \in \mathbb{N}_0^{N + 1}$ is the sequence of chunk boundaries and $\bm{a}=\bm{b}_{0:N-1}$ contains the positions of chunk ``anchor" frames, each of which is the very first frame of the chunk. After chunking, only these ``anchor" frames are retained, resulting in the dynamically decimated subsequence $\bm{y}_{\bm{a}} \in \mathbb{R}^{C \times N}$. 
Given segment boundaries we compute segment lengths $\bm{l} \in \mathbb{N}^N$ as
\begin{equation}
{l}_{m} = {b}_{m+1} - {b}_m, \quad m = 0, \ldots, N-1 %0 < m < N-1.
\end{equation}

\subsection{Dechunking}
\label{subsec:dechunking}

The dechunking procedure, given the anchor subsequence $\bm{y}_{\bm{a}}$ along with segment length information $\bm{l}$ employs $\mathcal{P}$ to expand back to original length $T$, 
\begin{equation}
    \hat{\bm{y}} \leftarrow \texttt{dechunk}\left(\bm{y}_{\bm{a}}, \bm{l} ; \mathcal{P}\right).
\end{equation}

Within each segment $m$ (with length $l_m > 1$), we perform one autoregressive predictor rollout for $l_m - 1$~steps, starting from the anchor frame $\bm{y}_{a_m}$. Concretely, letting $\hat{\bm{y}}_{t}\in\mathbb{R}^{C}$ denote the $t$-th latent frame in the prediction-based expanded sequence, we set
\begin{align}
\nonumber \hat{\bm{y}}_{a_m} &\leftarrow \bm{y}_{a_m}, \qquad \bm{h}_{0}\leftarrow \bm{h}_\text{init}, \\
\hat{\bm{y}}_{a_m\!+\!k}, \bm{h}_{k} &\leftarrow 
\mathcal{P}(\hat{\bm{y}}_{a_m\!+\!k\!-\!1}, \bm{h}_{k\!-\!1}\!),
\,\, k\!=\!1,\ldots,l_m\!-\!1,
\label{eq:dechunk_rollout}
\end{align}
where $\bm{h}_k$ represents the predictor's recurrent state, and $\bm{h}_\text{init}$ is a trainable initial state.
That is, each segment is represented by its anchor, and the remaining frames are synthesized via predictor rollout (with state reset at every segment boundary).

\subsection{Boundary selection}
\label{subsec:boundary_selection}

Chunking is guided by the assumption that latent \emph{predictability implies redundancy}: if a segment of latents can be approximated from its anchor via $\mathcal{P}$, then these frames can be dropped. Given a target number of anchors $N$ (kept length), we can then formulate the boundary selection problem as a constrained minimization problem of the overall approximation error. 
We provide two solvers: (i) an approximate but efficient greedy procedure, and (ii) an exact dynamic program (DP), similar in spirit to CodecSlime~\cite{wang2025codecslime} but with a predictability cost derived from our learned latent dynamics model.

\subsubsection{Efficient greedy algorithm}

Starting from the finest segmentation (every frame is a potential anchor), we compute one-step approximation errors for \emph{right-expanding} each segment using $\mathcal{P}$: for each boundary position, we measure the error incurred by replacing the right-neighboring anchor with the predictor output from the segment's current last frame. We then perform $T-N$ iterations, each time selecting the candidate expansion with the smallest incremental cost, executing the expansion (thereby removing one frame), and updating only the affected local cost for replacing the next right-neighboring frame.
Once a segment has been expanded, we set its anchor replacement cost to $+\infty$, so it can no longer be replaced. This rule ensures the invariance that once an expansion is committed, its local cost that was incurred is final and will not change in subsequent iterations. Thus, the final accumulated error decomposes as a sum of the chosen incremental costs.
A limitation of this freezing rule is that it can restrict the minimum achievable kept length $N$. %, as in a worst-case setting we could have $T/2$ segments with length 2 each, leaving no options for further compression. 
With a priority queue this runs in $O(T\log T)$ time after initializing local costs, enabling fast per-sample rate control at inference.

\subsubsection{Exact dynamic programming algorithm}

To formulate the DP solution, we start by defining a \emph{segment approximation cost}. For a candidate segment that starts at index $s$ and has length $\ell$, we roll out the predictor for $\ell-1$ steps starting from anchor $\bm{y}_s$ and measure cumulative error:
\begin{equation}
c(s,\ell) 
=
\sum_{i=1}^{\ell-1} 
d \left(\hat{\bm{y}}_{s+i}^{(s)}, \bm{y}_{s+i} \right),
\label{eq:segment_cost}
\end{equation}
where $\hat{\bm{y}}_{s+i}^{(s)}$ denotes the predictor rollout anchored at index $s$, and $d(\cdot,\cdot)$ is a latent-space distance (we use MSE in practice).

Let $\mathrm{DP}[t,j]$ be the minimum cost to cover the first $t$~timesteps using $j$~replacements. Then, if hyperparameter $K_{\max}$ is the maximum segment length,
\begin{equation}
\mathrm{DP}[t,j] =
\min_{1\le \ell \le K_{\max}}
\mathrm{DP}[t-\ell,j-\ell+1] + c(t-\ell,\ell),
\label{eq:dp_recurrence}
\end{equation}
with $\mathrm{DP}[0,0]=0$, $\mathrm{DP}[t,0]=+\infty$ for $t>0$ and $\mathrm{DP}[0,j]=+\infty$ for $j>0$. Backtracking recovers the selected boundaries. The time complexity is $O(T^2\,K_{\max})$ given precomputed costs $c(\cdot,\cdot)$. Notably, while rows must be processed in increasing 
$t$, the $j$-dimension allows parallel evaluation.

\subsection{Training}
\label{subsec:training}

We train the Re-Bottleneck components $\mathcal{R}_{\texttt{enc}}, \mathcal{R}_{\texttt{dec}}$ and predictor $\mathcal{P}$ while keeping the base autoencoder $\mathcal{A}$ frozen. Training targets two coupled objectives: (i) accurate multi-step latent prediction and (ii) preservation of reconstruction quality after chunking/dechunking.

To match open-loop usage, we train $\mathcal{P}$ with multi-step rollouts. Starting at every index $s \in [0, T - 2]$, we initialize $\hat{\bm{y}}^{(s)}_{s}\leftarrow \bm{y}_{s}$ and roll out Eq.~\eqref{eq:predictor_step} for $K_\text{roll}$ steps (chosen as a hyperparameter). We then define the rollout prediction loss as
\begin{equation}
\mathcal{L}_{\mathrm{roll}}
=
\frac{1}{\sum_{i=1}^{K_\text{roll}}(T-i)}
\sum_{i=1}^{K_\text{roll}}\sum_{s=0}^{T-i-1}
d_{\text{norm}}\left(\hat{\bm{y}}^{(s)}_{s+i},\ \bm y_{s+i}\right),
\end{equation}
where $d_{\text{norm}}$ is the target-standard-deviation-normalized MSE.

We additionally expose the model to the same chunking-dechunking transformation used during inference. For a sampled target kept fraction $\rho$ (or kept length $N=\lfloor\rho T\rfloor$), we compute boundaries $\bm{b}=\texttt{chunk}(\bm{y};N,\mathcal{P})$ \emph{without backpropagating through the discrete boundary selection} (i.e., treating $\bm{b}$ as constant), and dechunk to obtain $\hat{\bm{y}}$. On the frames replaced during this transformation, we compute an auxiliary prediction loss $\mathcal{L}_{\mathrm{valid}}$ using the same distance $d$. The full prediction loss is
$
\mathcal{L}_{\mathrm{pred}}
=
\lambda_\text{roll}
\mathcal{L}_{\mathrm{roll}}
+
\lambda_\text{valid}\mathcal{L}_{\mathrm{valid}},
$
and is used to train only the predictor $\mathcal{P}$ parameters.

We minimize a reconstruction loss through the Re-Bottleneck decoder:
\begin{equation}
\mathcal{L}_{\text{rec}}
=
d\left(\bm{z},\,\mathcal{R}_{\texttt{dec}}(\hat{\bm{y}})\right),
\label{eq:rec_loss}
\end{equation}
where $d$ is MSE. We additionally use an adversarial and a feature matching objective as done in~\cite{bralios2025re}.
This training encourages $\mathcal{P}$ to produce stable rollouts and the Re-Bottleneck reconstructions to remain compatible with the frozen autoencoder decoder, while enabling inference-time rate control via greedy/DP boundary selection.

\section{Experimental Setup}
\label{sec:setup}

In our experiments, we evaluate the reconstruction quality of our method across a range of operating points. We consider both greedy and DP-based chunking, and examine models trained at a fixed kept fraction $\rho$ (e.g., $\rho=0.5$) as well as \emph{scalable} models trained over a distribution of kept fractions, e.g., $\rho \sim \mathcal{U}(0.5, 0.995)$. All models are implemented as Re-Bottlenecks that augment the bottleneck of a frozen autoencoder $\mathcal{A}$.

\subsection{Baseline methods}
\label{sec:baselines}

We compare against methods for frame-rate reduction, which can be adaptive/non-adaptive and scalable/non-scalable. 
%To ensure a fair comparison, we implement and train all of them as Re-Bottleneck versions, using the same training data and model sizes as in our proposed methods.
For a fair comparison, we implement all baselines as Re-Bottleneck modules using the same base autoencoder, training data, and model capacity as our proposed methods, while allowing method-specific decimation components.

\begin{itemize}
\item \textbf{Conv-Downsample}. This baseline performs fixed-rate temporal downsampling using an additional Conv1d layer as the final encoder $\mathcal{R}_\texttt{enc}$ layer, and an additional transpose Conv1d as the first layer of the decoder $\mathcal{R}_\texttt{dec}$.
\item \textbf{CodecSlime}~\cite{wang2025codecslime}. We train a Re-Bottleneck in two stages, melt and cool. First, in the melt phase we train $\mathcal{R}_\texttt{enc}$, $\mathcal{R}_\texttt{dec}$ to be robust to segments created through frame-averaging. Then, in the cool phase, we fine-tune $\mathcal{R}_\texttt{dec}$ at $\rho=0.5$, using boundaries produced by the CodecSlime DP algorithm with a maximum segment length of $4$. Unlike Elastic Time, it does not utilize a learned predictor.
 \item \textbf{H-Net}~\cite{hwang2025dynamic}. A dynamic chunking layer proposed for text, whose boundary selection procedure is trained end-to-end. It is not scalable since the target compression ratio must be set as a loss hyperparameter during training.
\item \textbf{H-Net-YOTO}. To make H-Net scalable, we train a loss-conditional version following YOTO~\cite{Dosovitskiy2020You}. Both the encoder $\mathcal{R}_\texttt{enc}$ and decoder $\mathcal{R}_\texttt{dec}$ are conditioned on the target kept fraction $\rho$, which is sampled from a range. For both H-Net models during inference, we tune the boundary probability threshold to achieve the exact user-requested rate. 
\end{itemize}

\subsection{Model details}

As the frozen base autoencoder $\mathcal{A}$ we use the publicly available Stable Audio Open (SAO) VAE~\cite{evans2025stable}. It is a 156~M parameter model that compresses 44.1~kHz stereo audio into latents with 64 channels at a frame rate of 21.5~Hz.
Re-Bottleneck encoder $\mathcal{R}_\texttt{enc}$ and decoder $\mathcal{R}_\texttt{dec}$ are symmetric ConvNeXt-V2~\cite{woo2023convnext} networks, each with 6 blocks and hidden dimension 1024, resulting in 51~M additional parameters. Our predictor $\mathcal{P}$ has 0.5~M parameters and consists of 3 GRU layers followed by a SwiGLU-FFN~\cite{shazeer2020glu} block, all placed between linear projections to and from a 128~hidden dimension.   

\subsection{Training details}

\begin{figure*}[t]
    \centering
     \vspace{0pt}
        \centering
       \includegraphics[width=0.99 \linewidth]{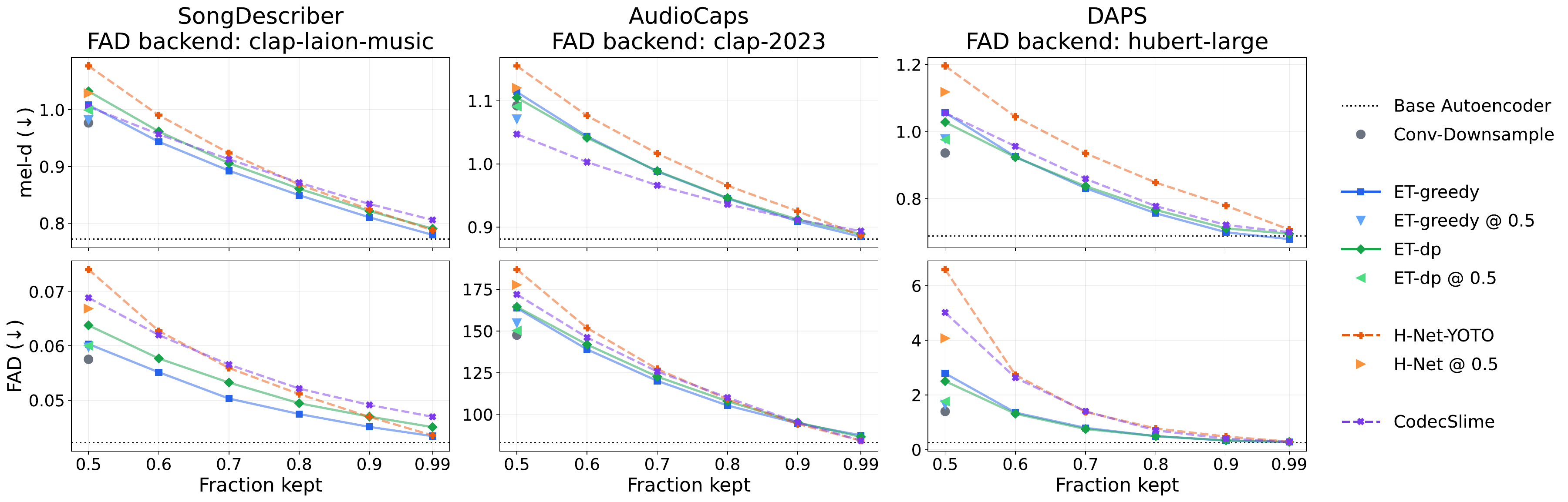}
       \caption{
       %Reconstruction quality across different compression levels. Top row: mel-spectrogram distance (mel-d, $\downarrow$). Bottom: FAD ($\downarrow$).
       Reconstruction quality as a function of latent frame-rate. The x-axis reports the kept fraction $\rho$ of the 21.5 Hz base latent frame-rate, ranging from 10.75 Hz at $\rho$=0.5 to 21.29 Hz at $\rho$=0.99.
       Top row: mel-spectrogram distance (mel-d, $\downarrow$). Bottom: FAD ($\downarrow$).}
    \label{fig:mel_fad}
    \vspace{-1.64em}
\end{figure*}

Our training data consists of a $4.8$k-hour mix of audio and music data, sampled at 44.1~kHz with most files being stereo. We use publicly available sources: AudioSet-balanced~\cite{gemmeke2017audio}, FSD50k~\cite{fonseca2021fsd50k}, BBCSoundEffects~\cite{bbc_sound_effects}, RWC~\cite{rwc}, MoisesDB~\cite{pereira2023moisesdb} and Jamendo-FMA-captions~\cite{lanzendorfer2025coarse, bogdanov2019mtg, defferrard2017fma}. To speed up training, we pre-encode all of our data, and train on latent chunks of size 96, corresponding to $\approx4.5$~s. We train for 250k steps with batch size 64, taking approximately 48~hours on a L40S GPU. We use AdamW~\cite{loshchilov2018decoupled} ($\text{lr}=10^{-4}$, $\text{wd}=10^{-4}$) for the discriminator, and a combined Muon~\cite{jordan2024muon, liu2025muon} ($\text{lr}=10^{-3}$, $\text{momentum}=0.95$) and AdamW ($\text{lr}=10^{-4}$, $\text{wd}=10^{-4}$) optimizer for the rest of the model. We chose $K_\text{max} = 12$ and $K_\text{roll} = 5$. %Hyperparameter $\lambda_{s,i}$ is set to $0.5$ for rollout positions used during dechunking and $0.4$ otherwise. 
We set $\lambda_\text{roll}=0.4$ and $\lambda_\text{valid}=0.5$.
The weight of the objective $\mathcal{L}_\text{rec}$ is set to $2.0$, while the adversarial objective is weighted by $0.5$ and the feature matching one by $1.0$. All models have KL-parameterized bottlenecks with a KL-objective weight of $0.0001$ like SAO. 

% TODO: Loss weights

\subsection{Evaluation \& Metrics}

We evaluate on unseen sets from diverse audio domains: instrumental music on SongDescriber~\cite{evans2025stable, manco2023song}, sound effects on AudioCaps (test)~\cite{kim2019audiocaps}, Chinese vocal music on MuChin~\cite{wang2024muchin}, and speech on DAPS~\cite{mysore2014can}. Each audio sample is trimmed to a random $10$~s chunk (consistent across eval runs).
Reconstruction quality is measured by SI-SDR, mel and STFT distances computed using \texttt{auraloss}~\cite{steinmetz2020auraloss}. We report averages over clips.
We measure distribution-level reconstruction quality by computing the Fréchet Audio Distance (FAD) using \texttt{fadtk}~\cite{gui2024adapting}.

\section{Results}
\label{sec:results}

We evaluate four main Elastic Time (ET) variants: two greedy-based models, one trained at a fixed kept fraction $\rho=0.5$ (denoted \textbf{ET-greedy@0.5}) and one trained over a range $\rho\sim \mathcal{U}(0.5,0.995)$ (\textbf{ET-greedy}), and two DP-based models, \textbf{ET-dp@0.5} ($\rho=0.5$) and \textbf{ET-dp} ($\rho\sim \mathcal{U}(0.5,0.995)$). In addition, we train \textbf{ET-dp-widerange}, with $\rho\sim \mathcal{U}(0.2,0.995)$ to test the impact of training for more aggressive compression.

Fig.~\ref{fig:mel_fad} reports mel-d and FAD for ET and the baselines described in Sec.~\ref{sec:baselines} across three datasets. 
Since each dataset corresponds to a different audio domain, we compute FAD using a domain-appropriate backbone; values are not directly comparable across columns.
We evaluate at kept ratios corresponding to per-sample average rates in the range $[10.75, 21.29]~\mathrm{Hz}$.

We first note that greedy and DP chunking perform similarly, suggesting that the efficient greedy procedure captures most of the benefit of the exact DP objective in this setting.
Next, H-Net baselines generally underperform across datasets. We hypothesize that this stems from a mismatch between the H-Net boundary-selection mechanism, originally designed for causal language modeling, and our finite-segment offline autoencoding setting. 
This observation supports the view that explicitly optimizing boundary selection, as in Elastic Time and CodecSlime, can ease the burden on the rest of the model by offloading temporal allocation decisions to a dedicated procedure.

Compared with CodecSlime, ET generally achieves lower mel-d and FAD on SongDescriber and DAPS, with the clearest gains in FAD. 
This suggests that the learned predictor provides a strong signal for boundary selection choices that better preserve perceptual quality under rate reduction. However, on AudioCaps, CodecSlime achieves slightly lower mel-d, while ET remains competitive in FAD. One plausible explanation is domain mismatch: around 82\% of our training data is music,
and the latent dynamics learned by $\mathcal{P}$ may generalize less effectively to the sound-effect content in AudioCaps.

Single-rate ET variants (ET-greedy @ 0.5 and ET-dp @ 0.5) improve over their scalable counterparts at $\rho = 0.5$, indicating a trade-off between operating-point specialization and rate scalability. 
We also find they close the gap to, and in some cases surpass, the fixed-rate Conv-Downsample baseline. 
That is notable since boundary-based compression introduces discrete boundary decisions instead of relying on a fixed-stride downsampling/upsampling pathway.
Our primary goal, however, is not to optimize solely for $\rho=0.5$, but to achieve {frame-rate scalability} while retaining competitive reconstruction quality. % while supporting user-controlled rate variation at inference time. 

Finally, Tab.~\ref{tab:recon_muchin_mel_stft_sisdr} extends the evaluation to MuChin and includes STFT-d and SI-SDR metrics. We note ET often improves over the dynamic baselines. ET-dp-widerange is competitive  at $\rho=0.5$, likely because training over a wider rate range makes it a moderate rather than extreme operating point. However, its slightly weaker performance at $\rho=0.9$ suggests a tradeoff.

\begin{table}[t]
\centering
\caption{
MuChin reconstruction metrics. Cells report mel-d~$\downarrow$ / STFT-d~$\downarrow$ / SI-SDR~$\uparrow$ (dB) at kept fractions $\rho \in \{0.5,0.7,0.9\}$.
}

\label{tab:recon_muchin_mel_stft_sisdr}
% \scriptsize
% \setlength{\tabcolsep}{4pt}
% \renewcommand{\arraystretch}{1.15}
\resizebox{\columnwidth}{!}{
\begin{tabular}{@{}l c c c@{}}
\toprule
\textbf{Model} & $\rho{=}0.5$ & $\rho{=}0.7$ & $\rho{=}0.9$ \\
\midrule
{ET-greedy @ 0.5}     & 1.11 / 1.71 / 2.3 &        &        \\
{ET-greedy}           & 1.14 / 1.73 / 1.6 & 1.03 / 1.65 / 3.4  & 0.95 / 1.58 / 4.6  \\
{ET-dp @ 0.5}         & 1.13 / 1.74 / 2.2 &        &        \\
{ET-dp}               & 1.16 / 1.75 / 1.6 & 1.05 / 1.66 / 3.4  & 0.96 / 1.59 / 4.7  \\
{ET-dp-widerange}     & 1.12 / 1.70 / 2.2 & 1.04 / 1.65 / 3.6  & 0.97 / 1.60 / 4.6  \\
\specialrule{.05em}{.1em}{.1em} 
{Conv-Downsample}     & 1.11 / 1.72 / 2.5 &       &        \\
\specialrule{.05em}{.1em}{.1em} 
{H-Net @ 0.5}         & 1.15 / 1.73 / 1.4 &        &        \\
{H-Net-YOTO}          & 1.20 / 1.80 / 0.3 & 1.05 / 1.66 / 2.9  & 0.96 / 1.59 / 4.4  \\
\specialrule{.05em}{.1em}{.1em} 
{CodecSlime}          & 1.14 / 1.73 / 0.7 & 1.04 / 1.65 / 2.6  & 0.96 / 1.59 / 4.3  \\
\bottomrule
\end{tabular}
}
\vspace{-1.8em}
\end{table}

\section{Conclusion}
\label{sec:conclusion}

We presented \textit{Elastic Time} (ET), a dynamic latent frame-rate mechanism for neural audio autoencoders that enables frame-rate scalable models via content-adaptive latent decimation. 
Implemented as a {Re-Bottleneck} plug-in on top of a frozen pretrained autoencoder, ET introduces a lightweight causal latent predictor for chunking and dechunking. 
Across multiple audio domains and compression levels, ET achieves competitive efficiency-quality tradeoffs relative to fixed-rate and prior dynamic chunking baselines, while enabling deployment-time rate control without external semantic supervision. 
More broadly, dynamic frame-rate scalability could improve the efficiency of downstream generation and long-context audio modeling.

\section{Acknowledgments}

% {\color{blue}Acknowledgments should be included only in the camera-ready version, not in the version submitted for review. For regular papers, pages 5 and 6, and for long papers, pages 9 and 10, are reserved exclusively for acknowledgments, disclosures of the use of generative AI tools, and references. No other content may appear on these pages. Any appendices must be contained within the first four pages for regular papers and within the first eight pages for long papers.

% Acknowledgments and references may begin on an earlier page if space permits.}

% \ifcameraready
%      The Interspeech 2026 organizers
% \else
%      The authors
% \fi
This work was supported by Electronics and Telecommunications Research Institute (ETRI) grant funded by the Korean government [26ZC1100, Development of Spatial Media Technology and Interaction Technology for Convergence of the Real and Virtual World].

\section{Generative AI Use Disclosure}
ChatGPT was used as an editing tool during the preparation of this manuscript.

\bibliographystyle{IEEEtran}
\bibliography{mybib}

\end{document}